\documentclass[]{agujournal2019}
\usepackage{amsmath, amssymb, lineno, soul, url, xcolor}

\journalname{Geophysical Research Letters}

\begin{document}

\title{Revised Estimates of Ocean Surface Drag in Strong Winds}
\authors{M. Curcic\affil{1} and B. K. Haus\affil{1}}

\affiliation{1}{Rosenstiel School of Marine and Atmospheric Science, University of Miami, 4600 Rickenbacker Causeway, Miami, FL 33149.}

\correspondingauthor{Milan Curcic}{mcurcic@miami.edu}

\begin{keypoints}
\item
New laboratory experiments confirm presence of air-sea drag saturation in strong winds
\item
Previous laboratory drag data found to be underestimated due to an error in scaling to 10-m height
\item
Corrected data show air-sea drag saturation at 34\% higher magnitude and 12\% lower wind speed
\end{keypoints}
\begin{abstract}

Air-sea drag governs the momentum transfer between the atmosphere and the
ocean, and remains largely unknown in hurricane winds.
We revisit the momentum budget and eddy-covariance methods to estimate the
surface drag coefficient in the laboratory.
Our drag estimates agree with field measurements in low-to-moderate winds,
and previous laboratory measurements in hurricane-force winds.
The drag coefficient saturates at $2.6\times10^{-3}$ and
$U_{10}\approx25\ m\ s^{-1}$, in agreement with previous laboratory results
by \citeA{takagaki12}.
During our analysis, we discovered an error
in the original source code used by \citeA{donelan04}.
We present the corrected data and describe the correction procedure.
Although the correction to the data does not change the key finding of 
drag saturation in strong winds, its magnitude and wind speed threshold 
are significantly changed.
Our findings emphasize the need for an updated 
and unified drag parameterization based on field and laboratory data.

\end{abstract}

\section*{Plain Language Summary}
We measure the rate of air-sea momentum transfer (surface drag) in strong winds
in a 15-m long wind-wave tank.
In support of previous work, we find further evidence that the drag saturates
(levels off) in hurricane-force winds, based on three different measurement
methods.
The level of drag saturation, however, is higher than previously thought.
The leading study that discovered the drag saturation in high winds had an error
in the source code used for the analysis of the data.
This error resulted in an overestimate of 10-m wind speed, and an underestimate
of the drag coefficient.
This finding is important because previous laboratory data that underestimate
the drag were used to implement surface flux parameterization in the most widely
used research and operational weather prediction model for tropical cyclone
applications.

\section{Introduction}

Air-sea drag determines the rate of vertical exchange of horizontal momentum
between the atmosphere and the ocean \cite{phillips66, large81, edson13}.
This exchange is mediated largely by ocean surface waves
\cite{donelan93, donelan97}, and in the most extreme weather conditions by
bubbles, spray, and spume 
\cite{andreas92, makin05, holthuijsen12}.
Correct formulation of the air-sea momentum flux as a surface boundary condition
is critical for numerical weather \cite{powers17, skamarock12, skamarock18} and
ocean prediction models \cite{shchepetkin05, chassignet07}.
This is especially true for the prediction of extreme weather events such as
hurricanes and winter storms \cite{chen16} where strong winds, high waves, and
surge compound to devastating effects on coastal communities.

Air-sea drag is challenging to measure in hurricanes because of their transient
and destructive nature.
To date, in situ measurements remain sparse and limited to tropical storm-force
winds ($U_{10}$ $<$ 27 m s$^{-1}$, \citeA{potter15}).
Other estimates of drag in the field are limited to indirect measurement
techniques, including the profile method using GPS dropsondes
\cite{powell03, holthuijsen12}, angular momentum budget \cite{bell12},
and upper-ocean momentum budget \cite{jarosz07, sanford11, hsu19}.
These estimates come with significant variability, possibly due to limited
sample sizes, indirect measurement methods, and drag dependence on wave-state.
Alternatively, drag can be measured in high winds in a controlled laboratory
environment 
\cite{donelan04, troitskaya12, takagaki12, takagaki16}.
Current consensus from laboratory and field data is that the drag coefficient
likely saturates in hurricane-force winds, and perhaps even decreases in more
extreme conditions \cite{holthuijsen12, soloviev14, donelan18}.

The main limitation of all laboratory drag estimates in high winds is
extremely short fetch, which caps the level of wave development that
can be achieved in wind-wave tanks.
Thus, these conditions are not equivalent to those in the field:
In the laboratory, hurricane-strength winds are forced over calm water,
and measurements are taken at fetch typically less than 10 m
\cite{donelan04, takagaki12};
in contrast, hurricane winds in the field are coupled with complex and mature
wave spectra that develop over fetches of order 100 km
\cite{wright01, donelan12, chen16, curcic16, collins18}.
Despite these limitations, the data collected by \citeA{donelan04} were
implemented as surface flux parameterization in the Weather Research and
Forecasting (WRF, \citeA{powers17}) model, and recommended for tropical cyclone
applications by \citeA{davis08}.
Specific values of drag measured in the laboratory have thus had direct
impact on weather and ocean prediction in both research and operational settings.

In this study, we re-evaluate the drag coefficient in high winds over initially
calm water similarly to \citeA{donelan04} and \citeA{takagaki12}.
We estimate the drag using the eddy-covariance flux method \cite{tennekes72, edson98}
and the momentum budget method \cite{donelan04} in 10-m wind speeds up to
50 m s$^{-1}$, in fresh and seawater.
Our main objective is to validate the momentum budget method and reproduce the
drag saturation results by \citeA{donelan04} in the same wind-wave tank as the
original study.
We repeat the experiment with more and newer instruments, in both fresh and
seawater, and calculate the error estimates for both eddy-covariance and
momentum budged methods.
During our data analysis, we identified a programming error in the original
source code by \citeA{donelan04}, which was used to scale the in situ wind speed
to a reference height of 10 m.
The error caused an overestimate of 10-m wind speed and an underestimate of
the drag coefficient.
We include the corrected drag coefficient values alongside our new results.
This study addresses the need to identify and better understand the
limitations of drag measurements in wind-wave flumes, and help interpret how
the laboratory results can be applied to air-sea processes in the field
and numerical models.

\section{The Momentum Budget in the Laboratory}

\begin{figure}[h]
  \centering
  \includegraphics[width=\textwidth]{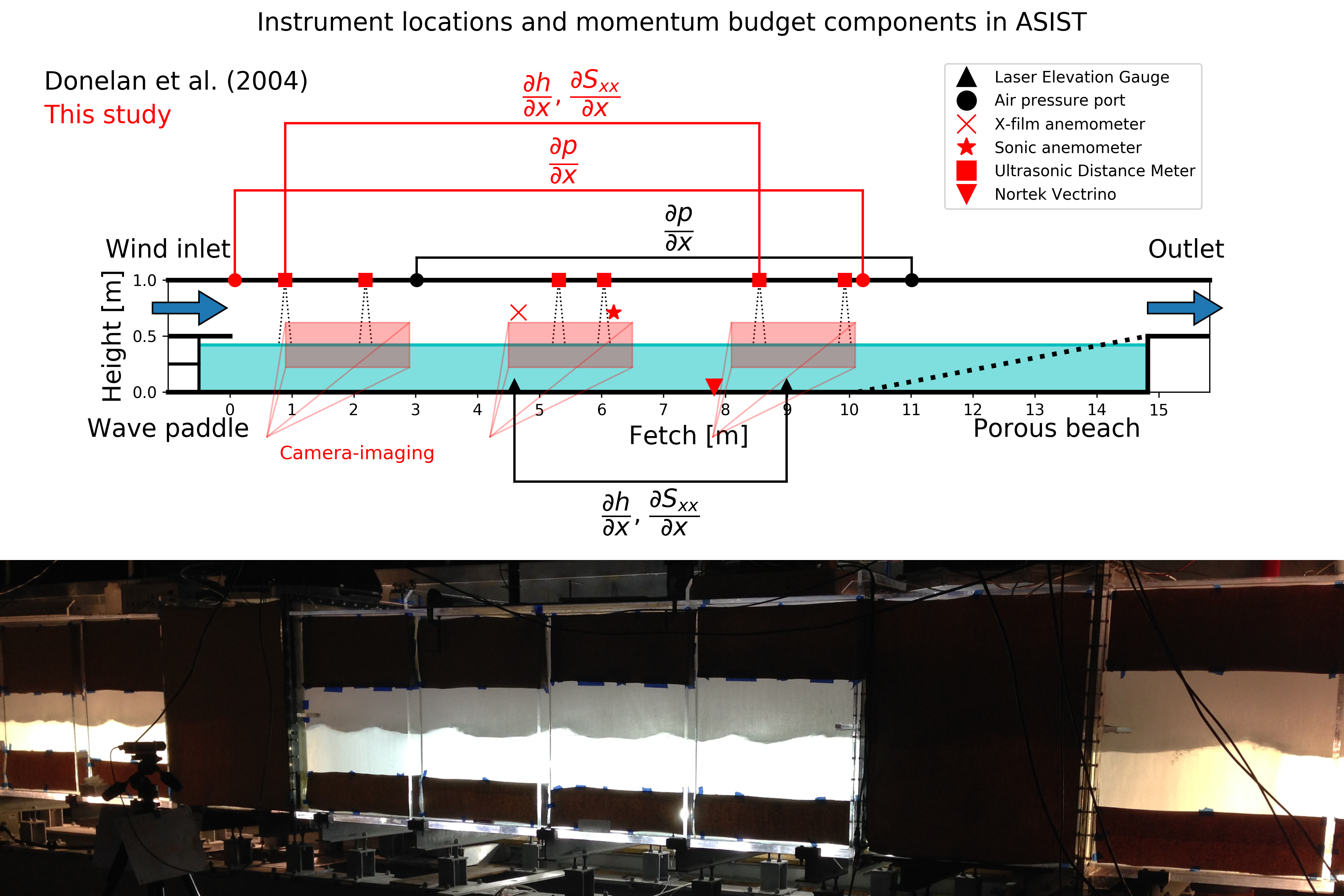}
  \caption{(Top) Laboratory set up and locations of instruments in the tank from
    \citeA{donelan04} (black) and this study (red). Connecting lines indicate
    the instrument positions used to compute pressure, elevation, and radiation
    stress gradients. (Bottom) Photograph of the ASIST wave tank during the
    experiment.}
  \label{fig01}
\end{figure}

We conducted the experiments in the Air-Sea Interaction Salt-Water Tank (ASIST)
at the University of Miami.
This tank has been successfully used to measure air-sea fluxes
\cite{donelan04, donelan18, jeong12}, wave modulation
\cite{donelan10, laxague17}, and sea-spray generation \cite{ortiz16}, among
others.
The tank is 15-m long with a 1 $\times$ 1 m cross-section.
The final 5 m of the tank feature a sloping porous beach designed to dissipate
waves while minimizing shoaling and reflection.
The laboratory set-up and the position of instruments are illustrated in Fig. \ref{fig01}.

We follow the momentum budget approach by \citeA{donelan04} to estimate drag in
high winds.
These conditions are prohibitive to the eddy-covariance method due to copious
spray that corrupts the measurement.
The momentum budget method in the tank is based on the steady-state momentum
balance between surface stress $\tau$,
mean slope of the water surface $\dfrac{\partial \eta}{\partial x}$,
radiation stress gradient $\dfrac{\partial S_{xx}}{\partial x}$,
air pressure gradient $\dfrac{\partial p_a}{\partial x}$,
and the bottom stress $\tau_b$:
\begin{equation}
\label{eq:momentum_budget}
\tau = \rho_w g h \dfrac{\partial \eta}{\partial x}
     + \dfrac{\partial S_{xx}}{\partial x}
     + h \dfrac{\partial p_a}{\partial x}
     + \tau_b
\end{equation}
where $\rho_w$ and $h$ are water density and mean water height,
respectively, $g$ is gravitational acceleration, $\eta$ is the
water elevation relative to the resting water height.
Specific locations in the tank over which the gradients are calculated for
both this and the reference study \cite{donelan04} are shown in Fig. \ref{fig01}.
Sampling the whole length of the tank is important because
$\partial p/\partial x$ and $\partial \eta/\partial x$ may vary in the
along-tank direction.
Thus, if we measure water elevation and air pressure in multiple locations
in the tank, and have an estimate of the bottom stress, then we can calculate
the surface stress using Eq. \ref{eq:momentum_budget}.
Drag coefficient at the reference height of 10 m is then $\tau / (\rho_a U_{10}^2)$,
where $\rho_a$ is air density and $U_{10}$ is wind speed at 10 m.

We measure the eddy-covariance momentum flux using a TSI hot-film anemometer
at 4.7 m fetch, and a Campbell Scientific IRGASON sonic anemometer at 6.1 m fetch.
We calibrated the hot-film anemometer using a pitot tube mounted at the same
fetch and height, which measured the horizontal component of the wind.
The air pressure gradient was measured via two small holes in the ceiling at
0.1 m and 10.2 m fetch.
The mean water elevation was measured using Senix ultrasonic distance-meters
mounted at the ceiling and looking down at the water surface.
The mean surface slope and radiation stress gradient were calculated by taking
the difference between the data at 8.6 m and 0.9 m fetch.
This yields a control section of 7.7 m, in contrast to 4.4 m in the momentum
budget set-up by \citeA{donelan04}.
In addition, we use high-speed cameras to film the water interface at three
2 m-long sections of the tank (0.9-2.9, 4.5-6.5, and 8.1-10.1 m fetch)
(Fig. \ref{fig01}, bottom).
We followed the image processing method by \citeA{laxague17} to extract the wave
frequency spectra and significant wave height as function of fetch.
Finally, we mounted a Nortek Vectrino at 7.8-m fetch to measure 3-dimensional
water velocity in the lower 4-cm profile, which provided an estimate of the
bottom stress $\tau_b$.
All measurements are averaged over 6-minute runs, following \citeA{donelan04}.
We propagate the instrument errors in our analysis to calculate the final error
bars for the drag coefficient.

We collected all data in 13 different wind conditions, from calm to
approximately 25 m s$^{-1}$ (centerline, $z = 0.29\ m$) in equal 6-minute
increments, in fresh and seawater.
The resting water depth was 0.42 m in all experiments.
This laboratory set-up enables: (a) direct comparison with previous laboratory
measurements \cite{donelan04, takagaki12} and observing the drag saturation;
and (b) testing the momentum budget, eddy-covariance, and side-camera imaging
techniques for application in larger wind-wave flumes.

\section{Drag Coefficient in High winds}

\begin{figure}[h]
  \centering
  \includegraphics[width=\textwidth]{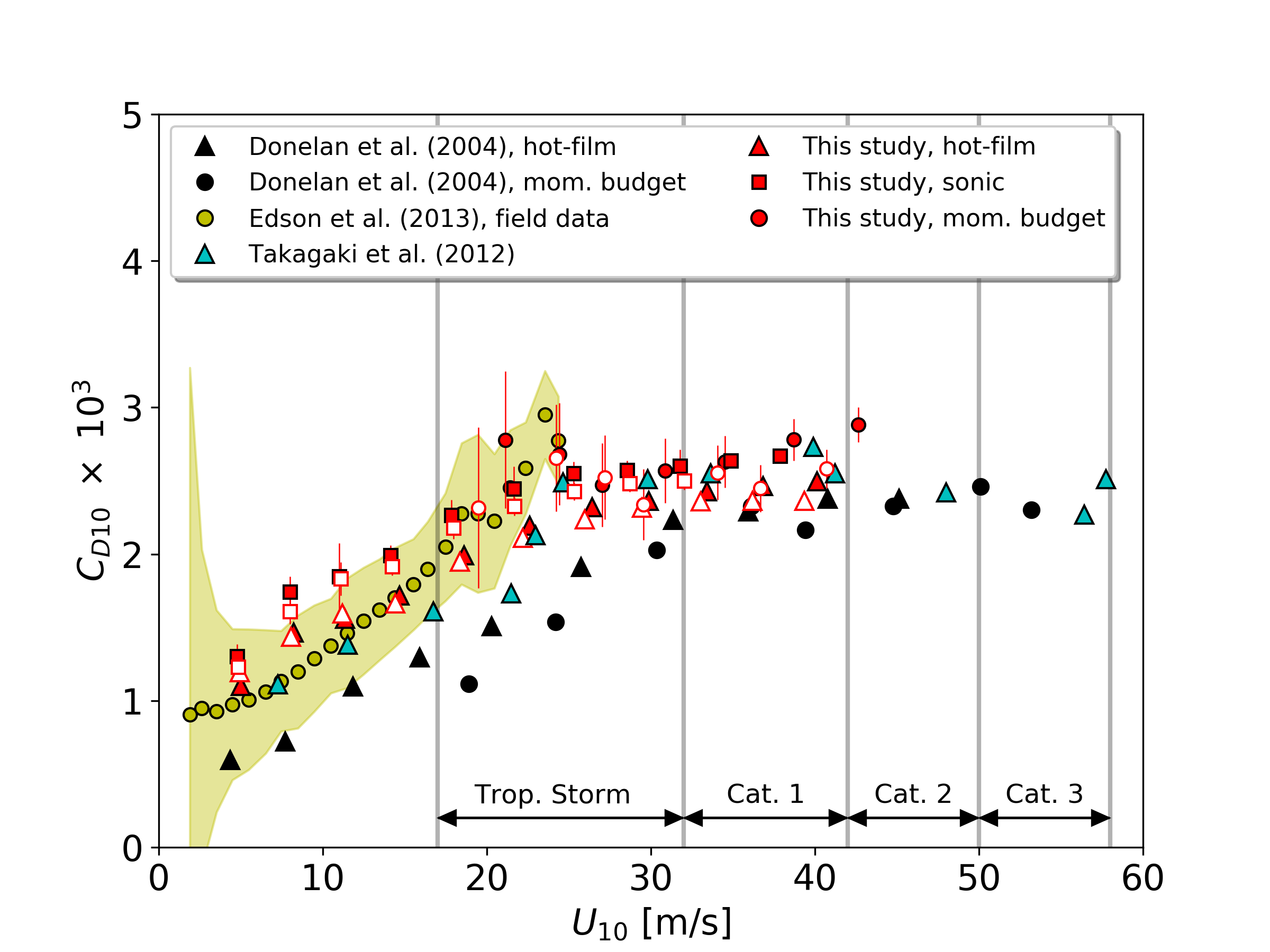}
  \caption{Drag coefficient scaled to 10-m height from eddy covariance and momentum budget
  from \citeA{donelan04} and this study, laboratory data from \citeA{takagaki12},
  and field measurements from \citeA{edson13}.
  Error bars from our study denote instrument error and/or method uncertainty.
  Field data are binned averages $\pm$ one standard deviation, and denote
  uncertainty from instrument errors and natural variability.
  Data from \citeA{donelan04} are shown in black.
  Solid and open markers from this study are from freshwater and seawater
  experiments, respectively.
  Vertical lines separate different Saffir-Simpson tropical cyclone intensity categories.}
  \label{fig02}
\end{figure}

We first evaluate the drag coefficient as function of wind speed, based on
measurements from the new experiments in the ASIST laboratory (Fig. \ref{fig02}).
We compute the drag coefficient from three sources: Hot-film and sonic
anemometers using the eddy-covariance method, and the momentum budget method.
To evaluate the robustness between separate experiments, we include both the
data from freshwater (solid markers) and seawater (open markers).
Unlike previous laboratory data, we calculate the error bars associated with
instrument accuracy and uncertainty of the momentum budget method.
For reference, we compare our results with the data from the field
\cite{edson13} and laboratory \cite{donelan04,takagaki12}.

There are three key takeaways from Fig. \ref{fig02}.
First, our new data confirm the presence of drag saturation in high winds.
The drag coefficient increases with wind speed up to $U_{10} \approx 25\ m\ s^{-1}$,
at a level of $C_D \approx 2.5\ \times 10^{-3}$, in agreement with the
laboratory data by \citeA{takagaki12}.
The transition wind speed of $25\ m\ s^{-1}$ 
is consistent with the results by \citeA{troitskaya12}.
Stress measured by the sonic anemometer (fetch of 6.1 m) is consistent with
that measured by \citeA{takagaki12} (fetch of 6 m), and both are slightly
higher relative to the stress measured with the hot-film anemometer (fetch of 4.7 m).
The drag values are expected to be sensitive to fetch because the roughness
elements are short wind-waves that grow downwind.
Main hypotheses that attempt to explain drag saturation include separation of
air-flow (sheltering) in the lee of steep wave crests \cite{donelan04}
and limited wave spectrum development due to wave breaking \cite{takagaki12, takagaki16}
and tearing of wave crests \cite{troitskaya12}.

Second, drag coefficients from three different sources in new experiments
are clustered together without notable outliers.
The new data also fall largely within the window of natural variability of
the field measurements in winds up to $25\ m\ s^{-1}$.
In high winds, the drag coefficient is saturated between $2.3\ \times 10^{-3}$
and $2.9\ \times 10^{-3}$, with an average value of $2.49\ \times 10^{-3}$.
The drag saturation signal is also consistent between fresh and seawater
experiments.
While the high-wind drag coefficient is slightly lower in seawater
($2.41\ \times 10^{-3}$) than in fresh water ($2.58\ \times 10^{-3}$),
we do not explore these effects further in this paper.
These results give us confidence in our instrumentation and the momentum budget
method for evaluating drag, and suggest that the drag saturation signal is
robust.
The momentum budget errors decrease with wind speed because
$\partial \eta / \partial x$ (along-tank slope) becomes significantly larger
than the measurement error of $\eta$.

Finally, while we reproduce the drag saturation found by \citeA{donelan04},
their drag estimates are consistently lower than the new laboratory data, as
well as field data \cite{edson13} in low-to-moderate winds.
In addition, their wind speed scaled to 10-m height extends considerably
further (highest value of $54\ m\ s^{-1}$) relative to this study (highest value
of $42\ m\ s^{-1}$).
We did not initially expect this discrepancy considering that both this study
and \citeA{donelan04} used the same wind-wave tank, resting water depth, and
wind forcing.
The following section describes the source of the discrepancy and the proposed
correction.

\section{Correction to Drag and Wind Data by Donelan et al. (2004)}

\begin{figure}[h]
  \centering
  \includegraphics[width=0.8\textwidth]{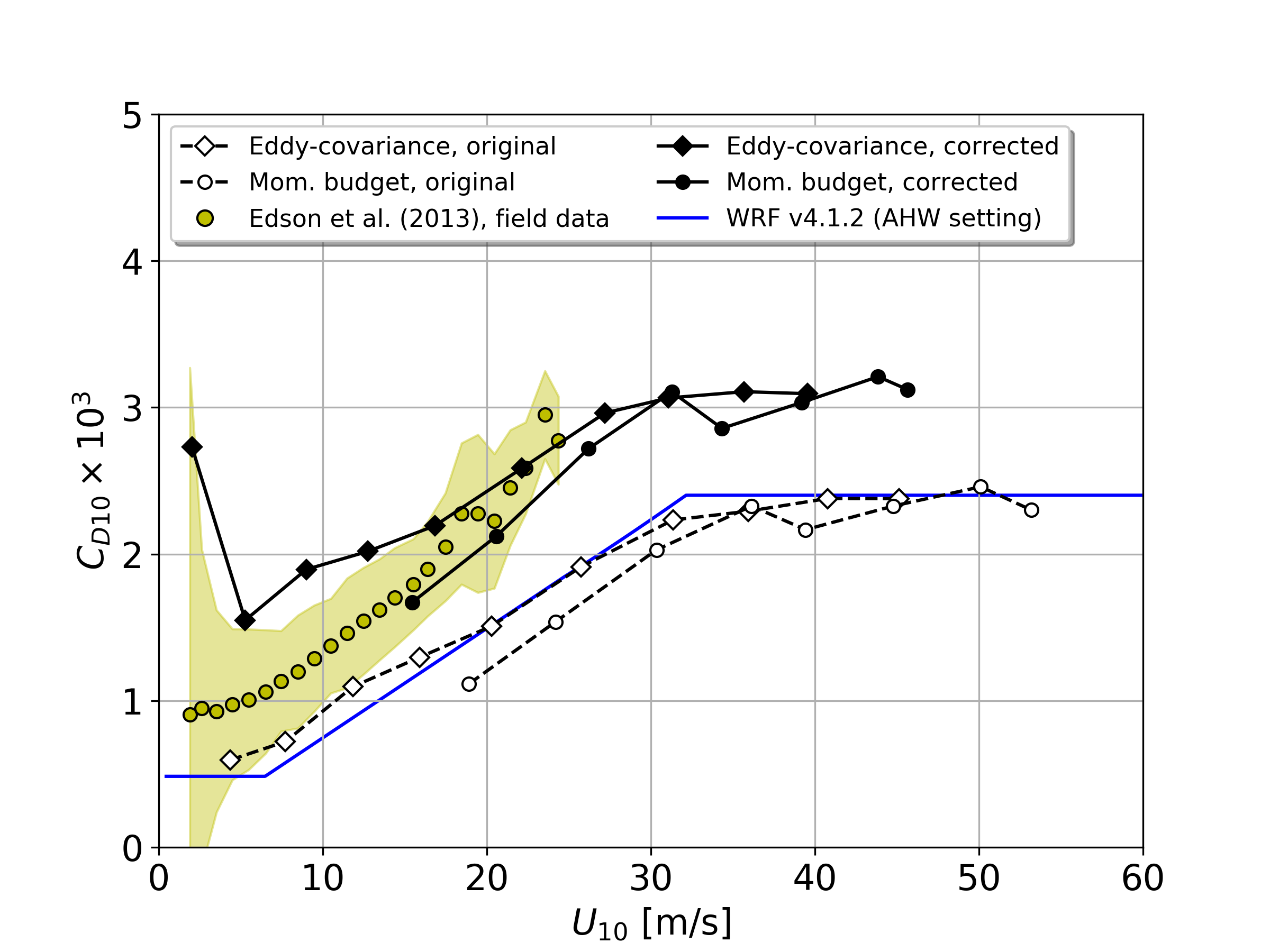}
  \caption{Original (black, dashed) and corrected (black, solid) drag coefficient data by \citeA{donelan04}.
	   Field data (mean $\pm$ one standard deviation) from \citeA{edson13} are shown in yellow.
     Surface drag parameterization in WRF (Advanced Hurricane WRF setting) is shown in blue.}
  \label{fig03}
\end{figure}

\begin{figure}[h]
  \centering
  \includegraphics[width=0.8\textwidth]{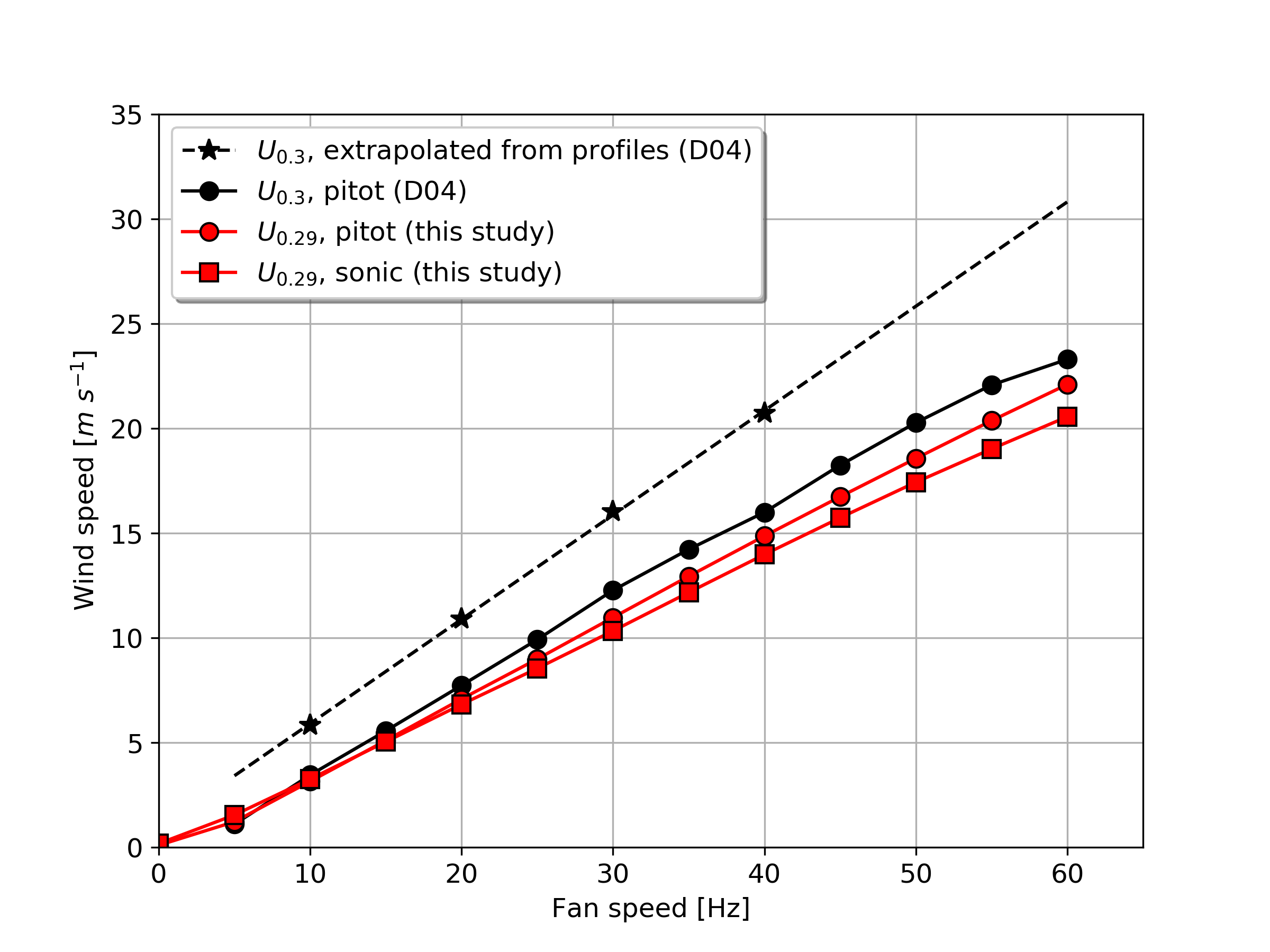}
  \caption{Extrapolated values (black stars) and in situ measurements
  (black circles) of $U_{0.3}$ from \citeA{donelan04}, and measured $U_{0.29}$
  by pitot (red circles) and sonic (red squares) anemometers during this study,
  as function of tank fan speed.}
  \label{fig04}
\end{figure}

\begin{figure}[h]
  \centering
  \includegraphics[width=\textwidth]{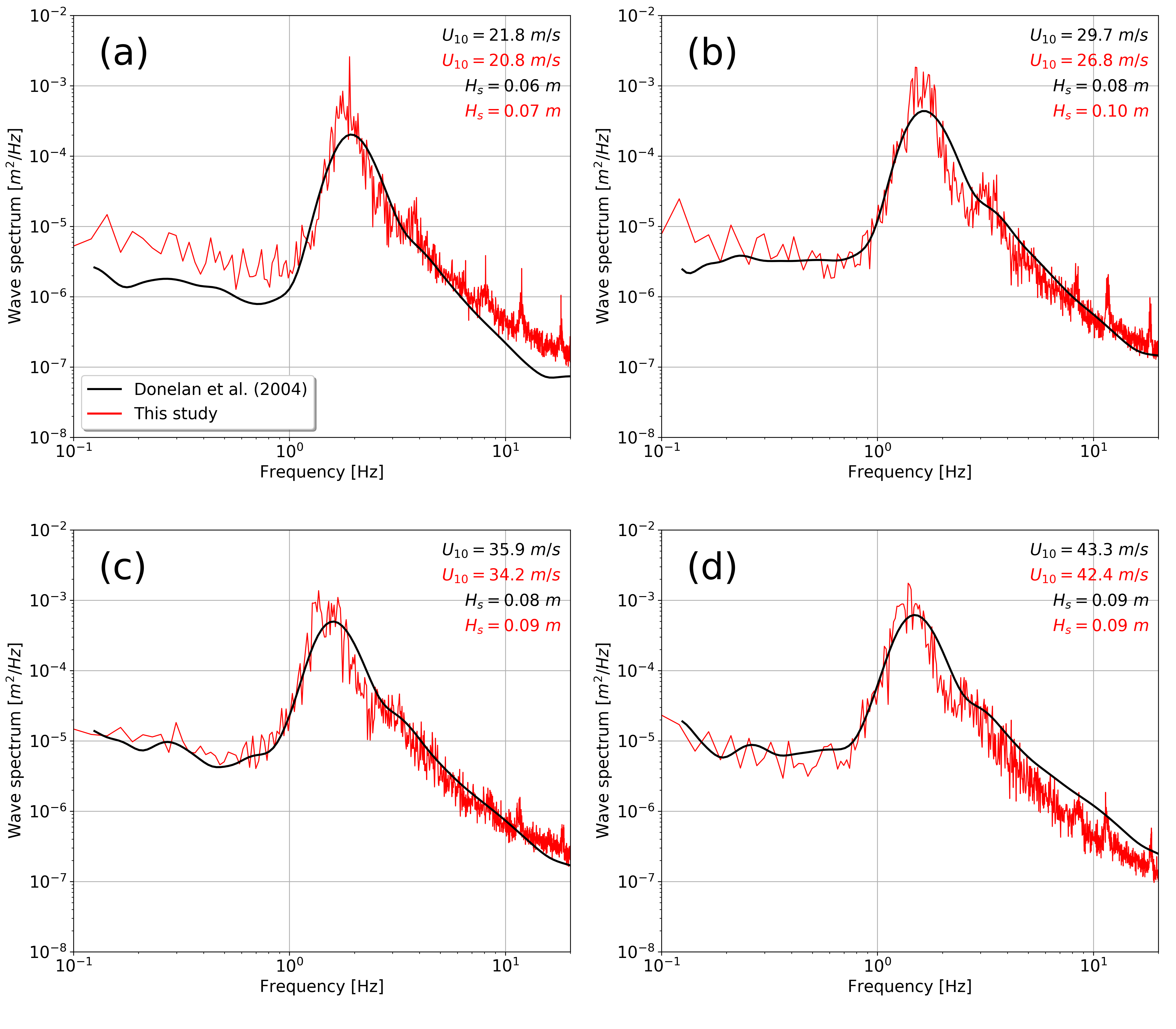}
  \caption{Measured wave elevation spectra at 9-m fetch in moderate-to-high wind
  conditions, from \citeA{donelan04} in black and this study in red.
  10-m wind and significant wave height values are shown in the upper-right
  corner of each panel.}
  \label{fig05}
\end{figure}

During the momentum budget analysis of the new laboratory data, we discovered
a critical flaw in the original source code used to calculate drag coefficient
values shown in Figure 2 of \citeA{donelan04}.
The correction is straightforward and can be applied using existing data from
the original experiment.
We show the original and corrected drag coefficients by \citeA{donelan04} in
Fig. \ref{fig03}.
Again, we show the field data by \citeA{edson13} for reference.
The correction brings the original drag coefficient saturation values from
$2.24\ \times 10^{-3}$ to $3.01\ \times 10^{-3}$, an increase of 34\%.
The saturation now occurs at lower wind speed, $U_{10} \approx 29\ m\ s^{-1}$
instead of $U_{10} \approx 33\ m\ s^{-1}$, a 12\% decrease, which is consistent
with both this study and \citeA{takagaki12}.
Corrected drag coefficients are more consistent with the field data in the
$15 - 23\ m\ s^{-1}$ range, as well as the new laboratory data from this study
(Fig. \ref{fig02}).
The WRF drag parameterization for the tropical flux configuration
\cite{davis08}, clearly modeled after the original laboratory data, inherited
the underestimate of the drag saturation magnitude, and overestimate of the
wind speed threshold at which the saturation occurs.

Where does the error in the drag coefficient by \citeA{donelan04} come from?
Wind and stress in the laboratory are measured at the height of the instrument,
in this case 0.3 m.
For convenience of comparison with prior work, field measurements, as well as
for application to models, in situ wind and drag coefficient are commonly scaled to some
reference height, typically 10 m above sea level for meteorological and
oceanographic applications.
Assuming neutral stability, the relationship is:

\begin{equation}
\label{eq:u10}
  U_{10} = U_z + \dfrac{u_{*}}{\kappa} log{\left( \dfrac{10}{z} \right)}
\end{equation}
where $U_z$ is the measured wind speed, $u_*$ is the friction velocity,
$\kappa$ is the Von K\'arm\'an constant (0.4), and $z$ is the height of the
measurement (0.3 m).
The drag coefficient scaled to 10-m height is then $\left(u_*/U_{10}\right)^2$.

The values used for $U_z$ in the original source code come from a different set
of measurements (wind and stress profiles) and do not correspond to the runs
during which the eddy-covariance and momentum budget data were collected.
Fig. \ref{fig04} shows that the wind speed extrapolated from profile
measurements is considerably higher than the in situ $U_{0.3}$, measured by
the pitot anemometer during the eddy-covariance and momentum budget runs.
We also show the in situ-measured wind speed at 0.29 m height ($U_{0.29}$)
from our study for comparison.
$U_{0.29}$ from this study is marginally smaller than the pitot-measured
$U_{0.3}$ from \citeA{donelan04}, primarily because the ASIST tank in 2003 was
configured to recirculate the air, allowing for slightly higher wind speeds,
and secondarily due to a small difference in measurement height.
The stress measurements are thus unaffected by the error.
The only aspects of the original study that are affected by the error are the
values of wind and drag coefficient scaled to 10-m height.
To correct the $U_{10}$ and $C_D$ estimates from eddy-covariance and
momentum budget approaches by \citeA{donelan04}, compute the friction velocity
$u_* = \sqrt{C_D} U_{10}$, and compute the new values of $U_{10}$ and $C_D$
following Eq. (\ref{eq:u10}), using in situ measured wind speed as $U_z$ instead
of that extrapolated from profile measurements.
This correction results in a decrease in $U_{10}$ and an increase in $C_D$,
as shown in Fig. \ref{fig03}.

The error and the corresponding correction for the first time explain both the
anomalously low values of $C_D$ relative to the field data, and the anomalously
high values of $U_{10}$ that we were not able to reproduce in this study.
The impact of this finding is tremendous because the data
in error were used to derive ocean surface drag parameterization in WRF, the
most widely used numerical weather prediction model \cite{davis08}.
This parameterization was later used by many
\cite{davis08a,davis10,nolan09,gopalakrishnan12,cavallo13,green13,green14}.
Further, the original data informed and influenced subsequent lines of research
\cite{takagaki12,takagaki16,chen07,chen13}, and was influential enough to be
covered by review papers \cite{black07,sullivan10}.
It is thus important to reflect back on the literature since the original
publication of the drag saturation data, with the new understanding that
the absolute magnitudes of drag coefficient have been underestimated.
However, the key finding of drag saturation by \citeA{donelan04} remains
and is further supported by this study.

As a further step of verification of stress and wave data between this study
and \citeA{donelan04}, we compare the power spectra of wave elevation
at 9-m fetch and four different levels of high wind speed (Fig. \ref{fig05}).
The spectral peak progressively shifts toward lower frequencies with higher
winds.
The sheltering of short waves is evidenced by a dip in the wave energy immediately
to the right of the windsea peak, between 2 and 3 Hz, likely due to the
air-flow separation in the lee of the dominant waves.
This process has been observed in the laboratory using particle-image
velocimetry \cite{reul99, buckley16, buckley19}, and was proposed by
\citeA{donelan04} as the dominant mechanism behind drag saturation.
At very high wind speeds, most of the momentum input occurs at the peak wave,
which dissipates and downshifts some of its energy, as evidenced by the
increase in wave energy for $f < 1\ Hz$ (Fig. \ref{fig05}b, c, d).
Long waves shelter shorter waves and limit their growth,
and at the same time dissipate due to direct impact from wind.
The former process acts to reduce the drag, the latter to increase it.
The spectra from \citeA{donelan04}, based on laser elevation gauge data,
are in close agreement with our own, which validates the high-speed camera
imaging method for non-intrusive measurement and analysis of wave spectra.

It is imperative to remember that the wind and wave conditions in the laboratory
are not equivalent to those in the open ocean, as discussed by \citeA{troitskaya12}.
Limitations include constrained water depth, fetch, directional spread
of wind-generated waves, as well as the air space between the water and the
ceiling of the tank.
Great care thus must be taken when designing experiments, collecting and
analyzing measurements, and interpreting the results.
Nevertheless, wind-wave laboratories provide an invaluable framework in which
the forcing conditions and 
the measurement instrumentation can be controlled at an unprecedented degree.
Such measurements are challenging, and often even impossible to collect in the
field.

\section{Conclusions}

We conclude with several key points and recommendations:

\begin{enumerate}

\item New laboratory data from the momentum budget method and eddy-covariance
flux measurements from two different instruments show further evidence of
air-sea drag saturation in strong winds, in both fresh and seawater.

\item The new drag estimates agree with the field observations \cite{edson13} in
low-to-moderate winds ($U_{10} < 25\ m\ s^{-1}$), and lab measurements
\cite{donelan04, troitskaya12, takagaki12} 
in high winds ($U_{10} > 25\ m\ s^{-1}$).

\item During the momentum budget analysis, we discovered a programming error in
the original source code used to generate drag estimates by \citeA{donelan04}.
The error caused the incorrect wind speed data array to be used when scaling
the wind speed and drag coefficient to a reference 10-m height.

\item Correction to the 10-m wind speed estimates from the dataset of
\citeA{donelan04} yields 10-m drag coefficients that agree better with field
data in low-to-moderate winds by \citeA{edson13}.
The revised drag saturates at $U_{10} \approx 29\ m\ s^{-1}$, at a level of
$C_D \approx 3.1 \times 10^{-3}$.
This saturation level corresponds to a 34\% increase relative to the original
estimates, with saturation occurring at 12\% lower wind speed threshold.

\item Laboratory measurements of drag in high winds over short fetch must be
cautiously compared to field measurements, due to the limitation of wind-wave
tanks to develop wave states comparable to those in the open ocean.
Drag parameterizations derived from such experiments
\cite{donelan04,troitskaya12,takagaki12,takagaki16}, 
as well as this study, should be used
mindfully and cautiously when implemented into numerical weather prediction
models such as WRF \cite{powers17} or MPAS \cite{skamarock12,skamarock18}.

\item This work illustrates that not only are weather and ocean prediction models
vulnerable to programmer errors, but that the datasets and analyses from which
parameterization schemes are derived may contain errors as well.
There is thus an increasing level of urgency to adopt open data, open software,
and open science practices across the theoretical, observational,
and numerical modeling disciplines.

\item Further work is needed to bridge the gap between the wind and wave
conditions in the open ocean and those that can be recreated in the laboratory.
The next step in this line of work will yield wind, wave, and drag measurements
in the 23-m long SUSTAIN wave flume at the University of Miami, capable of
generating extreme-force (Category 5 hurricane) winds.
These experiments will also explore the role of background wave conditions on
drag, such as complex spectra that are found in the open ocean.

\item Finally, a revised, consensus drag coefficient parameterization based on 
up-to-date field and laboratory measurements is needed for implementation 
in the current generation of uncoupled weather and ocean prediction models.

\end{enumerate}

\acknowledgments
We thank all members of the SUSTAIN laboratory at the University of Miami who
helped with the set up and calibration of instruments and data acquisition.
Discussions with William Drennan, Nathan Laxague, and Andrew Smith helped
improve this study.
Two anonymous reviewers helped improve the manuscript.
The data and code to reproduce the figures in this manuscript can be found at
https://zenodo.org/record/3731659.
The original data and code by \citeA{donelan04} can be found at
https://zenodo.org/record/3731657.
This work was supported by the National Science Foundation grant number 1745384.

\bibliography{references}

\end{document}